\documentclass[review]{elsarticle}
\usepackage{graphicx,float}
\usepackage{subcaption}
\begin{document}

\author[1]{ Jacob D. Baxley\corref{cor1}}
\ead{jacob.baxley351@topper.wku.edu}
\author[1]{David Lambert}
\author[2]{Mauro Bologna}
\author[1]{Bruce J. West}
\author[1]{Paolo Grigolini}

\cortext[cor1]{Corresponding author}
\address[1]{Center for Nonlinear Science, University of North Texas, P.O. Box 311427, Denton, Texas 76203-1427}
\address[2]{Departamento de Ingenier\'ia El\'ectrica-Electr\'onica, Universidad de Tarapac\'a, Arica, Chile}

\title{Unveiling Pseudo-Crucial Events in Noise-Induced Phase Transitions}


\begin{abstract}
Noise-induced phase transitions are common in various complex systems, from physics to biology. In this article, we investigate the emergence of crucial events in noise-induced phase transition processes and their potential significance for understanding complexity in such systems. We utilize the first-passage time technique and coordinate transformations to study the dynamics of the system and identify crucial events. Furthermore, we employ Diffusion Entropy Analysis, a powerful statistical tool, to characterize the complexity of the system and quantify the information content of the identified events. Our results show that the emergence of crucial events is closely related to the complexity of the system and can provide insight into its behavior. This approach may have applications in diverse fields, such as climate modeling, financial markets, and biological systems, where understanding the emergence of crucial events is of great importance.
\end{abstract}
\begin{keyword}
self-organization, noise-induced phase transition, waiting-time distribution, particle swarm optimization, Diffusion Entropy Analysis, pseudo-signs, inverse power-law index, multiplicative fluctuations, crucial events.
\end{keyword}
\begin{highlights}
    \item Noise-induced phase transitions exhibit pseudo-crucial events that are key to understanding complexity.
\item Diffusion entropy analysis can be used to accurately detect (pseudo-)crucial events in time-series data.
\item Demonstrates how to supplement multiplicative fluctuations to generate crucial events in noise-induced phase transitions, providing a more accurate detection of self-organization processes.
\end{highlights}
\maketitle
\section{Introduction}
The study of complexity in physics and biology has revealed the emergence of self-organization in a variety of natural phenomena \cite{selforganization,microtubules}. Self-organization is believed to be responsible for crucial events (CEs)\cite{garland}, which correspond to sudden renewals in the biological dynamics\cite{rejuvenation}. The time between neighboring CEs is described by a waiting-time probability distribution function (PDF) with an inverse power-law (IPL) index $\mu$ ranging from $1$ to $3$. Developing an algorithmic procedure for time-series analysis that can accurately detect these processes of self-organization is of great importance.

This paper presents an analysis technique for detecting pseudo-CEs associated with self-organization. Initially motivated by a perceived connection with the growth rate of developing fish, the project evolved into an exploration of the statistical properties of a surrogate time series generated by the stochastic logistic equation with multiplicative noise, also known as the Schenzle and Brand model. We investigate a form of phase transition called a noise-induced phase transition, which was first discovered by Schenzle and Brand \cite{schenzlebrand}. Although this phenomenon has generated considerable interest, there has been limited understanding of the dynamics at play.

We show that the multiplicative form adopted by Schenzle and Brand does not yield genuine CEs at the phase transition. Instead, we demonstrate how to supplement multiplicative fluctuations to generate pseudo-CEs. In Section II, we review the results of Schenzle and Brand, focusing on the equilibrium generated by the adoption of the Stratonovich prescription. In Section III, we present theoretical arguments for the origin of the IPL index $\mu = 1.5$ observed in the time series generated by the Schenzle and Brand model. Using the first passage time (FPT) technique, we derive $\mu = 1.5$ in both the Stratonovich and Ito prescriptions, as detailed in Section IV.

To explain why the observed events are not genuine CEs, we discuss the physics of the Schenzle and Brand model using a coordinate transformation that reveals the diffusion process underlying the events. Numerical results supporting our theoretical arguments are presented in Section VI. In Section VII, we provide additional theoretical arguments to demonstrate that CEs are incompatible with the fact that the fluctuation $\xi$ is entirely random.

In the concluding Section VIII, we suggest future work related to the possible manifestation of biological intelligence through the phenomenon of pseudo-crucial events in cancer growth. We also highlight the potential of particle swarm optimization as a tool for further exploring the dynamics of self-organizing systems.

\section{The Stochastic Logistic Equation}

In this section, we study a stochastic version of the logistic equation, given by:

\begin{equation} \label{xrepresentation}
\dot{x} = \alpha x - \beta x^2 + x \xi,
\end{equation}

where $x$ is the population of a species, $\alpha$ and $\beta$ are positive constants, and $\xi$ is a Gaussian white noise process with zero mean and intensity $q$ (i.e., $\langle \xi(t) \rangle = 0$ and $\langle \xi(t) \xi(s) \rangle = 2q \delta(t-s)$). The term $x \xi$ represents environmental fluctuations that affect the growth rate of the population.

The probabilistic picture explaining the time evolution of the PDF $p(x,t)$ of the population density is based on the parameter $q$, which is related to the intensity of the noise.

The equation of motion for $p(x,t)$ if we adopt the Ito prescription is given by:

\begin{equation}\label{fp1}
\frac{\partial}{\partial t}p(x,t)=-\frac{\partial}{\partial x}\left[\alpha x-\beta x^2\right]p(x,t)+q \frac{\partial^2}{\partial x^2}x^2 p(x,t),
\end{equation}

or, if we adopt the Stratonovich prescription, by

\begin{equation}\label{ffp2}
\frac{\partial}{\partial t}p(x,t)=-\frac{\partial}{\partial x}\left[\alpha x-\beta x^2\right]p(x,t)+q \frac{\partial}{\partial x}x \frac{\partial}{\partial x}x p(x,t).
\end{equation}

Note that Eq. (\ref{ffp2}) is equivalent to the reorganized version:

\begin{equation}\label{fp1bis}
\frac{\partial}{\partial t}p(x,t)=-\frac{\partial}{\partial x}\left[\left(\alpha+ q\right) x-\beta x^2\right]p(x,t)+q \frac{\partial^2}{\partial x^2}x^2 p(x,t).
\end{equation}

Schenzle and Brand \cite{schenzlebrand} showed that Eq. (\ref{ffp2}) yields the equilibrium PDF:

\begin{equation}\label{right}
p_{\mathrm{eq}}(x)=\frac{\left(\frac{\beta}{q}\right)^{\frac{\alpha}{q}}}{\Gamma\left(\frac{\alpha}{q} \right)}
x^{\frac{\alpha}{q}-1}
\exp\left[-\frac{\beta x}{q}\right],
\end{equation}

where $\Gamma(x)$ is the Gamma function. This PDF is a power law for small $x$ and has an exponential cutoff for large $x$, and its shape is determined by the dimensionless parameter $\frac{q}{\alpha}$.

Our focus is on understanding the behavior of this system, specifically on identifying the role of the randomness parameter $q$ in the system's dynamics. We show that the parameter $q$ plays a crucial role in determining the stability of the system, and that it can induce a noise-induced phase transition. We further demonstrate that this transition leads to pseudo-CEs where the system undergoes instantaneous renewal, resulting in an IPL PDF of waiting times with an index $\mu=3/2$. We also investigate the effect of noise correlations on the system's dynamics and show that they can significantly modify the behavior of the system. Our results have important implications for understanding the dynamics of real-world systems subject to environmental fluctuations, such as biological populations or financial markets.

\section{Complexity and Waiting-Time PDFs}

Blinking quantum dots exhibit intermittent fluorescence where the time durations of light and darkness have a waiting-time PDF with an IPL index $\mu$\cite{kuno}. Tang and Marcus proposed a theory that leads to 
\begin{equation}
\mu = \frac{3}{2}. 
\end{equation}
This value of $\mu$ appears to be a universal property of diffusion processes generated by random fluctuations of the form 
\begin{equation} \label{diffusion}
\dot x = \xi,
\end{equation}
where the noise $\xi$ has a scaling $\delta$ such that $x(t)\propto t^\delta$.

Assuming that the regression of $x(t)$ to an earlier value $x_0$ is renewal, the waiting-time PDF of the time distance between neighboring crossings of the same value $x_0$ is an IPL with index $\mu$ given by 
\begin{equation}\label{propagation}
\mu = 2 - \delta.
\end{equation}
This equation was used by Failla {et al} \cite{failla} to discuss the Kardar-Parisi-Zhang (KPZ) theory\cite{kpz}. The formula was derived under the renewal assumption to connect the IPL index $\mu$ to the scaling $\delta$ of the fluctuation $\xi$ generated by the dynamical process under study.

To evaluate the complexity of a time series $\xi$, researchers in the field of complexity convert the time series $\xi$ into the diffusion process $x(t)$ and evaluate its scaling $\delta$. It is commonly assumed that $\delta = \frac{1}{2}$ signals lack complexity, but this assumption is not quite correct. For instance, if $\mu = \frac{3}{2}$ is a consequence of $\delta = \frac{1}{2}$, as we prove in this paper in the case of noise-induced phase transition, the conclusion that there is no complexity in the process is correct. However, the evaluation of the scaling $\delta$ is a delicate problem that requires the use of the method of Diffusion Entropy Analysis (DEA).

In 2001, DEA was adopted to detect the scaling generated by CEs\cite{giacomo}. The authors of \cite{giacomo} found that the scaling evaluated by DEA depends on the rules adopted by the walker to move at the occurrence of CEs. In the special case when the walker makes always a jump ahead at the occurrence of a CE, 
\begin{equation} \label{jumpahead}
    \delta = \mu - 1.
\end{equation}
In this case, $\mu = \frac{3}{2}$ yields $\delta = \frac{1}{2}$, affording the wrong impression that the process under study is not complex.

This is not a reasonable conclusion, as we can prove studying the Dunbar effect\cite{dunbar}. The authors of Ref.\cite{dunbar} developed a theoretical approach to Dunbar's well-known observation that $N = 150$ is the optimal number of connections that the brains of primates, including human beings, can establish with the members of their society. This theoretical work established that the maximal scaling $\delta = 0.7$ is realized with $N=150$, moving from $\delta = \frac{1}{2}$. Since the statistical analysis of Ref.\cite{dunbar} was done using the jumping-ahead rule\cite{giacomo}, it is plausible that in this case, $\mu = \frac{3}{2}$, and that this is a genuine signal of CEs.

It is worth noting that the scaling of $\delta = \frac{1}{3}$ in KPZ is a result of the IPL index generated by the $x$-re-crossings being the same as the IPL index of $\xi$. As a consequence, the IPL index for KPZ is $\mu = \frac{5}{3} = 1.6666$, which is equivalent to the Dunbar IPL index when $N= 150$ established by the walking ahead rule. Thus, the statistical analysis in Ref. \cite{failla} that showed a transition from $\delta = 0.5$ to $\delta = 0.7$ indicates a change in the complexity of a system at criticality.

\section{Fokker-Planck Equation and First-Passage Time}\label{FPTSection}
We will use Weiss's first-passage time (FPT) technique \cite{weiss} to analyze the Ito and Stratonovich representations.

\subsection{Ito representation}

In this subsection we consider a partial differential equation that describes the evolution of a PDF $p(x,t)$ with respect to time $t$ and space $x$. The equation is given by:
\begin{equation}
    \frac{\partial }{\partial t}p(x,t)=-\frac{\partial }{\partial x} \left[\alpha x-\beta x^2\right]p(x,t)+q \frac{\partial^2 }{\partial x^2}x^2 p(x,t)
\end{equation}
where $\alpha$, $\beta$, and $q$ are constants. This equation describes a stochastic process, which is a mathematical model for a system that involves random events. In this case, the process describes the population of a biological system, where the population is subject to random fluctuations due to environmental factors. The equation is known as the Fokker-Planck equation, and it describes the PDF of the particle as it evolves in time.

The current density $J(x,t)$ is given by:
\begin{equation}
    J(x,t)=- \left[\alpha x-\beta x^2\right]p(x,t)+q \frac{\partial }{\partial x}x^2 p(x,t).
\end{equation}
The equilibrium PDF $p_{eq}(x)$ is found by setting $J(x,\infty)=0$, which leads to:
\begin{equation}
    p_{eq}(x)=\frac{\left(\frac{\beta }{q}\right)^{\alpha/q -1}}{\Gamma (\alpha/q -1)} x^{\frac{\alpha }{q}-2} \exp\left[-\frac{\beta x}{q}\right]
\end{equation}
where $\Gamma$ is the gamma function. To ensure the integrability of $p_{eq}(x)$, we require that $\alpha/q>1$. The equilibrium PDF represents the long-term behavior of the stochastic process, where the population has reached a steady state.

To analyze the behavior of the system, we take the Laplace transform of the Fokker-Planck equation. This leads to a differential equation in the Laplace domain, which can be solved to obtain the Laplace transform of the PDF. The solution is given by:
\begin{eqnarray}\nonumber
&&\tilde{p}(x,s)=\exp\left[-\frac{\beta  x}{q}\right] x^{\frac{\sqrt{(q-\alpha )^2+4 q s}+\alpha -3 q}{2 q}}
\\\nonumber
&&\left(c_1 U\left[\frac{\sqrt{(q-\alpha )^2+4 q s}+q-\alpha}{2 q},\frac{\sqrt{(q-\alpha )^2+4 q s}+q}{q},\frac{\beta  x}{q}\right]+\right.
\\ \label{anal1}
&&\left. c_2L\left[-\frac{\sqrt{(q-\alpha )^2+4 q s}+q-\alpha}{2 q},\frac{\sqrt{(q-\alpha )^2+4 q s}}{q},\frac{\beta  x}{q}\right]\right)
\end{eqnarray}
where $s$ is the Laplace variable, $U$ is the confluent hypergeometric function, $L$ is the Laguerre function, and $c_1$ and $c_2$ are constants. To determine these constants, we need to impose boundary conditions on the solution, which include continuity, discontinuity of the current density, and normalization.

Next, we focus on the solution for $x<x_0$ and impose the integrability of the probability at $x=0$, which requires setting $c_1=0$. Then, for $0<x<x_0$, we obtain the following expression for $\tilde{p}(x,s)$:
\begin{eqnarray}\nonumber
&&\tilde{p}(x,s)=\exp\left[-\frac{\beta  x}{q}\right]c(s,x_0) x^{\frac{\sqrt{(q-\alpha )^2+4 q s}+\alpha -3 q}{2 q}}
\\ \label{anal2}
&&L\left[-\frac{\sqrt{(q-\alpha )^2+4 q s}+q-\alpha}{2 q},\frac{\sqrt{(q-\alpha )^2+4 q s}}{q},\frac{\beta  x}{q}\right].
\end{eqnarray}
We then consider the case where $\beta\ll q$, which simplifies the expression for $\tilde{p}(x,s)$ further. In this case, we obtain:
\begin{eqnarray}\nonumber
&&\tilde{p}(x,s)\approx c(s,x_0) x^{\frac{\sqrt{(q-\alpha )^2+4 q s}+\alpha -3 q}{2 q}}
\\ \label{anal3}
&&L\left[-\frac{\sqrt{(q-\alpha )^2+4 q s}+q-\alpha}{2 q},\frac{\sqrt{(q-\alpha )^2+4 q s}}{q},0\right]
\end{eqnarray}
and the FPT function $f_{FPT}(s)$ from $x$ to $y$ is approximately:
\begin{eqnarray}\label{fpt1}
f_{FPT}(s) \approx \left(\frac{ x}{y}\right)^{\frac{\sqrt{(q-\alpha )^2+4 q s}+\alpha -3 q}{2 q}}.
\end{eqnarray}
Finally, we invert the Laplace transform to obtain the FPT function $f_{FPT}(t)$ for $x<y<x_0$ and $t\to \infty$:
\begin{equation}\label{fpt2}
f_{FPT}(t)\approx c(x,y)\frac{\exp\left[-\frac{1}{4} (\frac{\alpha}{q}-1)^2 q t-\frac{\log ^2\left(\frac{x}{y}\right)}{4 q t}\right]}{ \sqrt{ t^3}}
\approx c(x,y)\frac{\exp\left[-\frac{1}{4} (\frac{\alpha}{q}-1)^2 q t\right]}{ \sqrt{ t^3}}.
\end{equation}
This expression is valid when $x<y<x_0$ and in the limit of large $t$.
\subsection{Stratonovich representation for FPT PDF}
In this subsection, we introduce the Stratonovich representation and present the corresponding results for the FPT PDF. The Stratonovich representation is an alternative way of defining stochastic integrals that differs from the more commonly used Ito representation. In this case, we use Eq. (5) to derive the equilibrium PDF, which is given by:

\begin{eqnarray}\label{fpt1_5}
 	f_{FPT}(t)\approx c(x,y)\frac{\exp\left[-\frac{1}{4} \left(\frac{\alpha}{q}\right)^2 q t-\frac{\log ^2\left(\frac{x}{y}\right)}{4 q t}\right]}{ \sqrt{ t^3}}
 	\approx c(x,y)\frac{\exp\left[-\frac{1}{4} \left(\frac{\alpha}{q}\right)^2 q t\right]}{ \sqrt{ t^3}}.
 \end{eqnarray}

To obtain the FPT PDF, we repeat the same calculations as those done for the Ito case. The resulting expression is given by Eq. (\ref{fpt1_5}). This expression describes the PDF for the FPT from point $x$ to point $y$ when $x<y<x_0$ and as $t\to \infty$.

We note that the FPT PDF has a Gaussian decay with the square root of time in the denominator. However, the decay is faster (for $a>q$) than that obtained using the Ito representation. The FPT PDF depends on the parameter $c(x,y)$, which represents the probability that a population with initial population $x$ will reach a population of $y$ before hitting the absorbing boundary at $x_0$.

The expression for the FPT PDF, given by Eq. (\ref{fpt1_5}), reveals that it is highly sensitive to the initial conditions. In particular, the PDF depends on the relative distance between $x$ and $y$, as well as the ratio of the noise intensity to the potential strength. When the noise intensity is much smaller than the potential strength, the FPT PDF simplifies to the expression given by the second term in Eq. (\ref{fpt1_5}).

Our results provide new insights into the behavior of stochastic processes subject to external noise and potential fields. These insights can be used to better understand a wide range of physical systems, including the diffusion of molecules in biological cells, the motion of particles in turbulent fluids, and the behavior of financial markets.

\subsection{Preliminary Remarks on the FPT Approach}

Here we provide some preliminary remarks on the  FPT approach and its implications. We note that the adoption of this approach can sometimes yield results that conflict with our intuition. Specifically, using the Ito approach leads to an IPL PDF at the critical point of the Stratonovich approach.

To further illustrate this point, we examine Eq. (\ref{fpt2}), which shows that the exponential truncation disappears at $q = \alpha$ in the Ito prescription. However, the analytical calculations are done using the Ito prescription, which can result in some discrepancies.

We now turn their attention to the first-passage time calculation using the Stratonovich prescription. This calculation yields Eq. (\ref{fpt1_5}), which is characterized by an exponential truncation that is not cancelled at $q = \alpha$, as predicted by the Stratonovich prescription.

Overall, these preliminary remarks highlight the complexities involved in using the  FPT approach and the importance of carefully considering the choice of prescription.

\section{Coordinate Transformation}
 
Continuing from the last section, the Stratonovich representation provides an alternative approach to studying the stochastic process described by Eq. (\ref{xrepresentation}). The transformation from the Ito interpretation to the Stratonovich interpretation involves expressing the equation in terms of the variables 
 \begin{equation}
 y \equiv \ln \left(\frac{\beta}{\alpha}x\right)\quad\textrm{and}\quad\tau\equiv\alpha t.
 \end{equation}
 and $\tau = \alpha t$, which yields the equation 
 \begin{equation} \label{additive}
\frac{dy}{d\tau}\equiv\dot y = 1 - e^{y} + \frac{\xi}{\sqrt{\alpha}}.
\end{equation}
This transformation replaces the multiplicative fluctuation $x\xi(t)$ with the additive fluctuation $\frac{\xi}{\sqrt{\alpha}}$, which simplifies the analysis. However, it also introduces a non-polynomial potential 
\begin{equation}
V(y) \equiv e^{y} -y,
\end{equation}
which attracts our attention to negative values of $y$.

Using the Stratonovich interpretation, the equation of motion becomes 
\begin{equation} \label{additive2}
\dot y = - \frac{d V}{dy} + \sigma\xi,
\end{equation}
where $V(y)$ is the potential introduced by the transformation. The associated Fokker-Planck equation is given by  
\begin{equation} \label{fp}
 \frac{\partial}{\partial \tau} p(y,\tau) = \left(\frac{\partial}{\partial y} \left(\frac{d}{dy}V(y)\right) + \frac{q}{\alpha} \frac{\partial^2}{\partial y^2} \right) p(y,\tau),
\end{equation}
which yields the equilibrium PDF $p_{eq}(y) = \frac{e^{-\frac{\alpha V(y)}{q}}}{Z}$, where $Z$ is a normalization factor.

Notably, the equilibrium PDF $p_{eq}(x)$ in the original Ito representation can be obtained by transforming $p_{eq}(y)$ to $p_{eq}(x)$ using $p_{eq}(x)dx = p_{eq}(y(x))\left|\frac{dy}{dx}\right|dx$, where $y(x) = \ln\left(\frac{\beta}{\alpha}x\right)$. This yields an equilibrium PDF identical to the one derived earlier, as given by Eq. (\ref{right}). Numerical integration of the equation of motion in the Ito representation using the logarithmic transformation $y = \ln x$ leads to a $y$-equilibrium that can also be expressed in terms of the original variables $x$ and $t$, as given by Eq. (\ref{right}).

To assess whether the index $\mu = 3/2$ is indicative of CEs, we adopt the Stratonovich representation for the numerical analysis.
\section{Numerical integration and analysis}
To obtain the results presented in this paper, we numerically integrated Eq.~(\ref{xrepresentation}) using an Eulerian integration scheme. For the Ito case, we used the following numerical approximation:

\begin{equation}
x(t+1)=x(t)+\alpha x(t)-\beta x^2(t) +x(t) \sqrt{2D}\xi(t),
\end{equation}

where $\xi(t)$ is a delta-function correlated Gaussian noise with zero mean and unit variance, and the integration time step is set to 1. In the Stratonovich case, we used the following numerical approximation:

\begin{equation}
x(t+1)=x(t)+(\alpha-D) x(t)-\beta x^2(t) + x(t)\sqrt{2D}\xi(t).
\end{equation}

These numerical prescriptions are consistent with those described in \cite{Mannella2002}. The simulations and analysis were performed in Python, using the logistic equation with parameters $\alpha = 0.01$, $\beta=0.001$, and $q = 0.0099$. The code used for this analysis is available on GitHub at https://github.com/JacobBaxley/LogisticWork.

\begin{figure}
  \includegraphics[width=\linewidth]{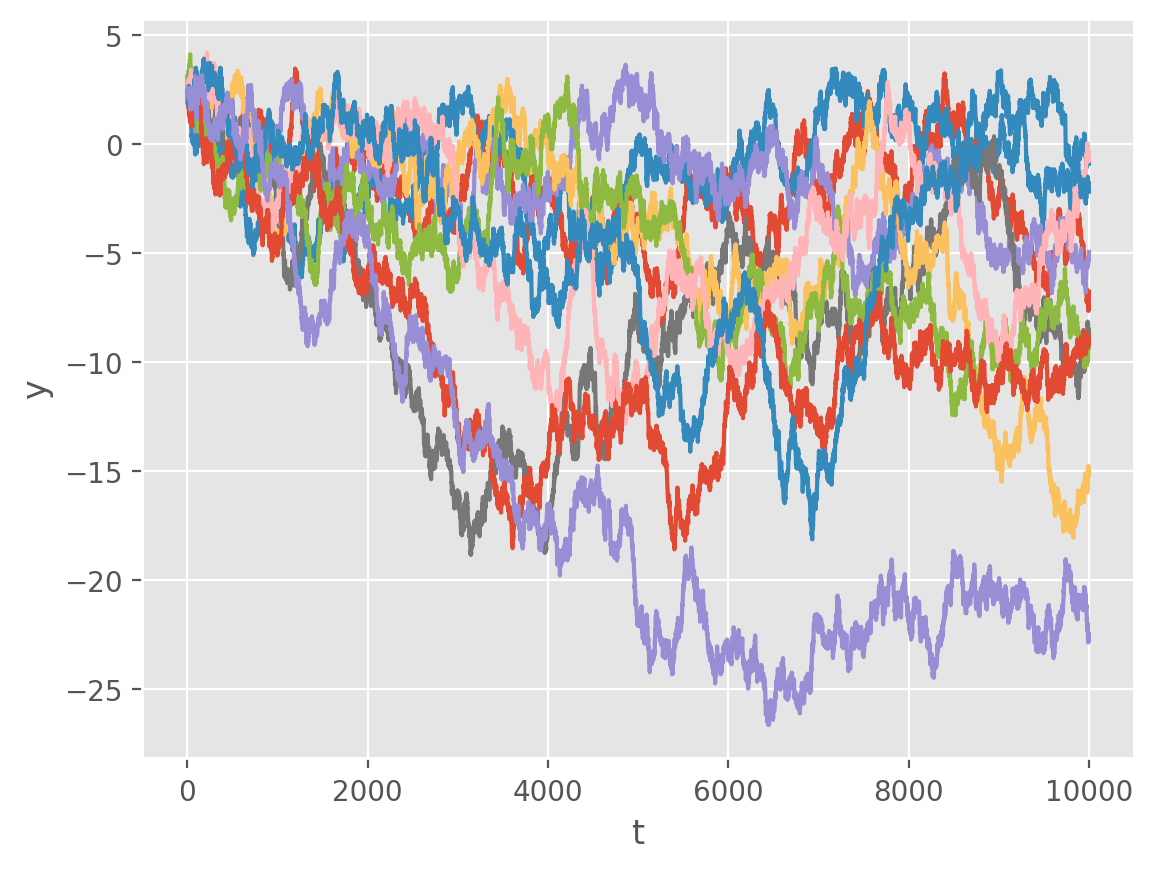}
  \caption{Ensemble of realizations of trajectories generated by the logistic equation.}
  \label{fig:logistic_ensemble}
\end{figure}

In particular, we analyzed ensembles of logistic equations, as shown in Figure \ref{fig:logistic_ensemble}. Our numerical approach allowed us to study the behavior of the system in detail and investigate the effect of different noise types and integration schemes. We emphasize that the small values of $\alpha,$ $\beta,$ and $D$ used in our simulations are necessary to ensure the stability and accuracy of the numerical integration scheme.
\subsection{DEA of Noise-Induced Phase Transitions in the Logistic Equation}
DEA is a powerful method that has been previously utilized to study various phenomena, including the social phenomenon of teen births \cite{scafetta}. In our previous work, we used a modified version of DEA called MDEA with stripes to study the scaling of CEs in noise-induced phase transition processes \cite{giacomo, memorybeyondmemory}. CEs can be hidden in a sea of additional fluctuations, and the adoption of stripes helps in their detection.

In a recent study \cite{garland}, it was shown that Fractional Brownian Motion (FBM), despite being a source of anomalous diffusion, should not be confused with a source of CEs. Here, we apply DEA with stripes to investigate whether noise-induced phase transition processes generate extended diffusion processes. We find that these diffusion processes are driven by totally random fluctuations and generate recrossings of the $x$-thresholds at neighboring times with a waiting-time PDF characterized by an IPL with $\mu = 3/2$, as explained in Section \ref{FPTSection}.

\begin{figure}[ht!]
  \centering
  \begin{subfigure}[b]{0.4\linewidth}
    \includegraphics[width=\linewidth]{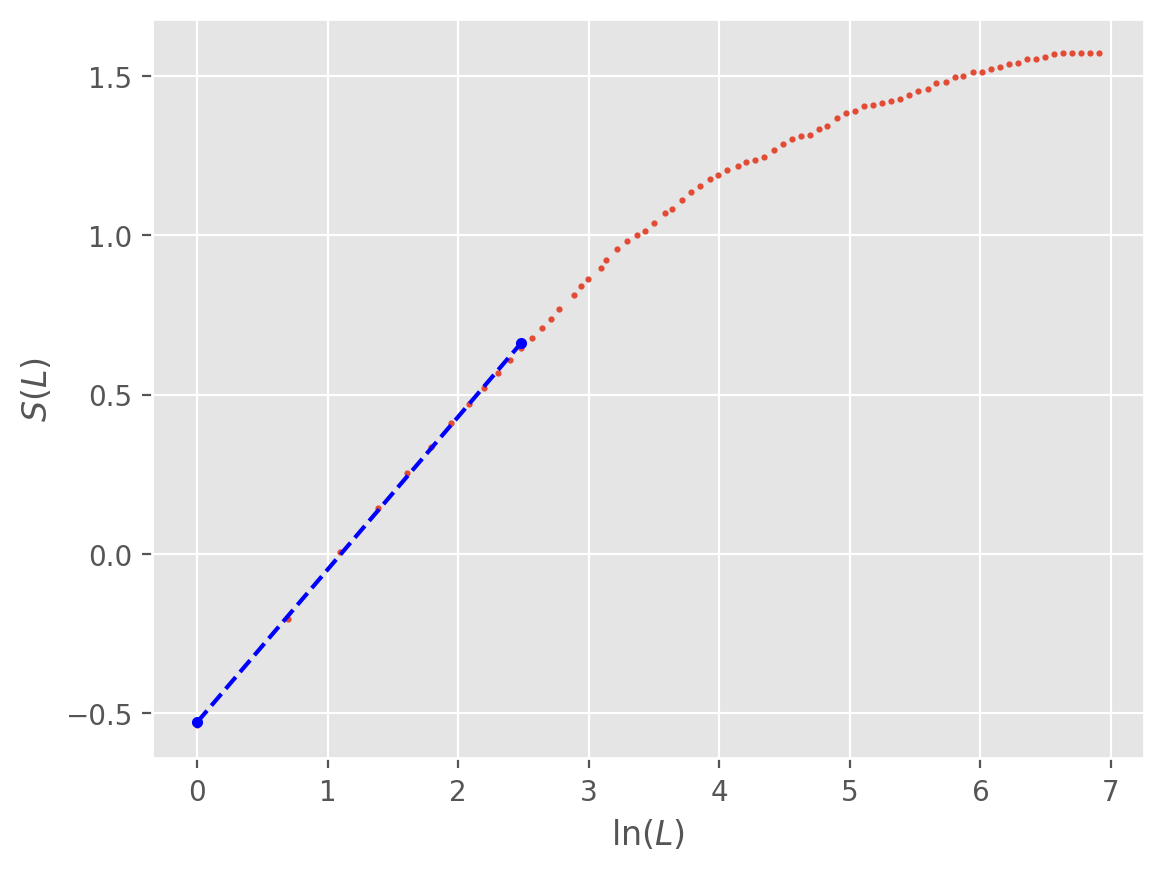}
    \caption{ODEA Stratonovich: $ \delta = 0.48 \pm 0.004$}
  \end{subfigure}
  \begin{subfigure}[b]{0.4\linewidth}
    \includegraphics[width=\linewidth]{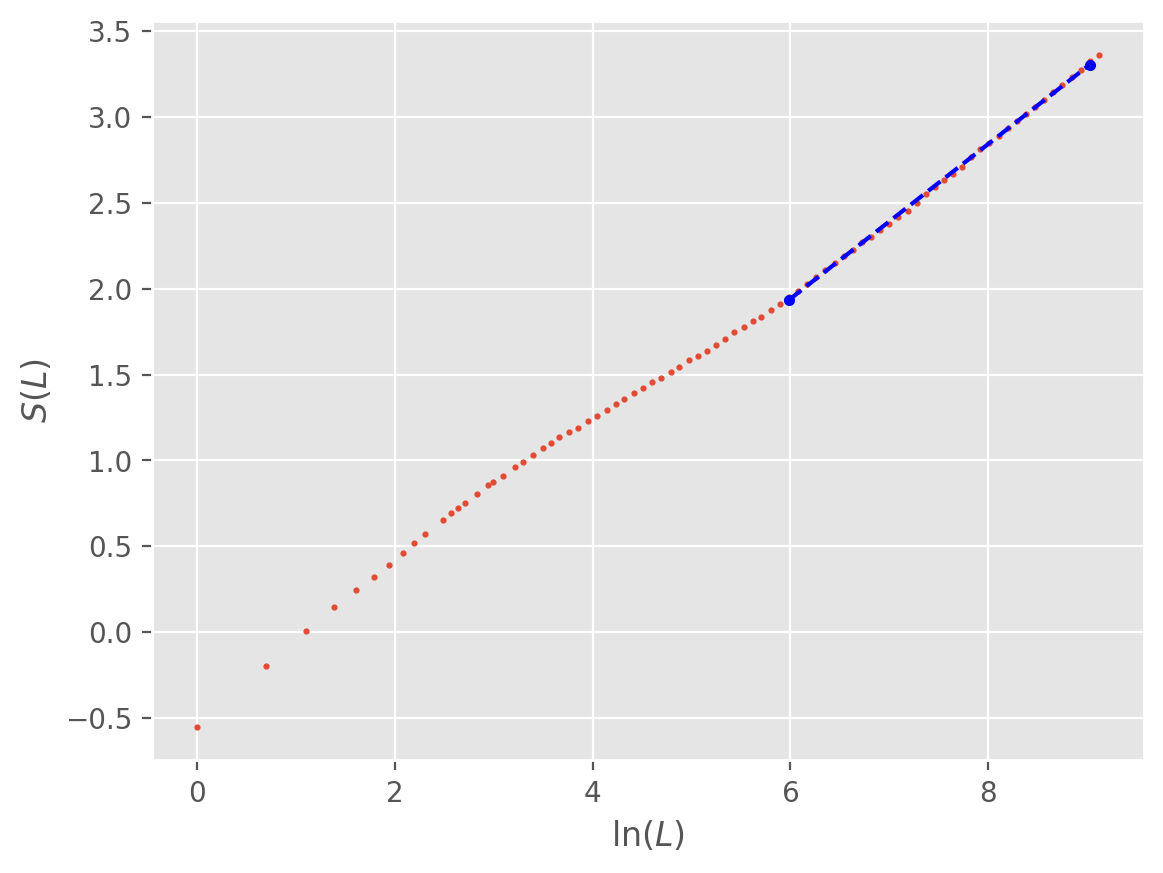}
    \caption{ODEA Ito: $\delta = 0.45 \pm 0.002$}
  \end{subfigure}
  \begin{subfigure}[b]{0.4\linewidth}
    \includegraphics[width=\linewidth]{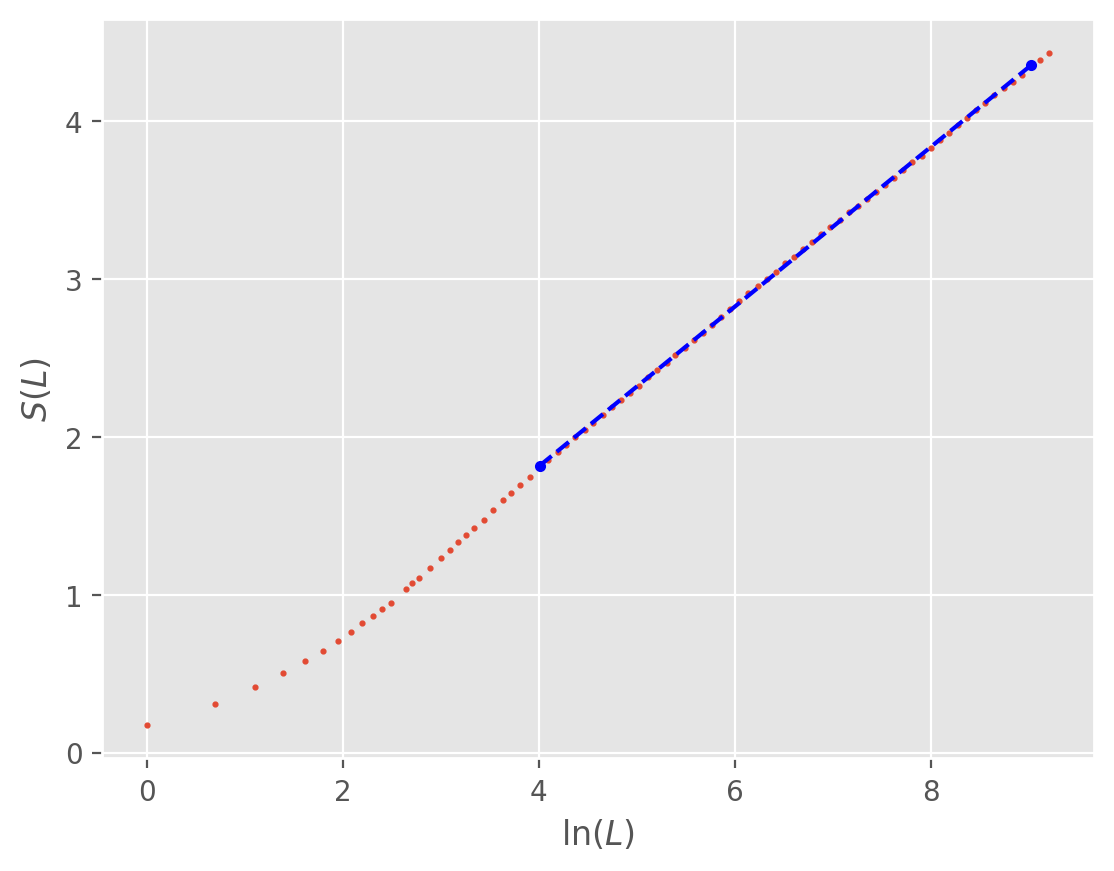}
    \caption{DEA Stratonovich: $ \delta = 0.5 \pm 0.007$}
  \end{subfigure}
  \begin{subfigure}[b]{0.4\linewidth}
    \includegraphics[width=\linewidth]{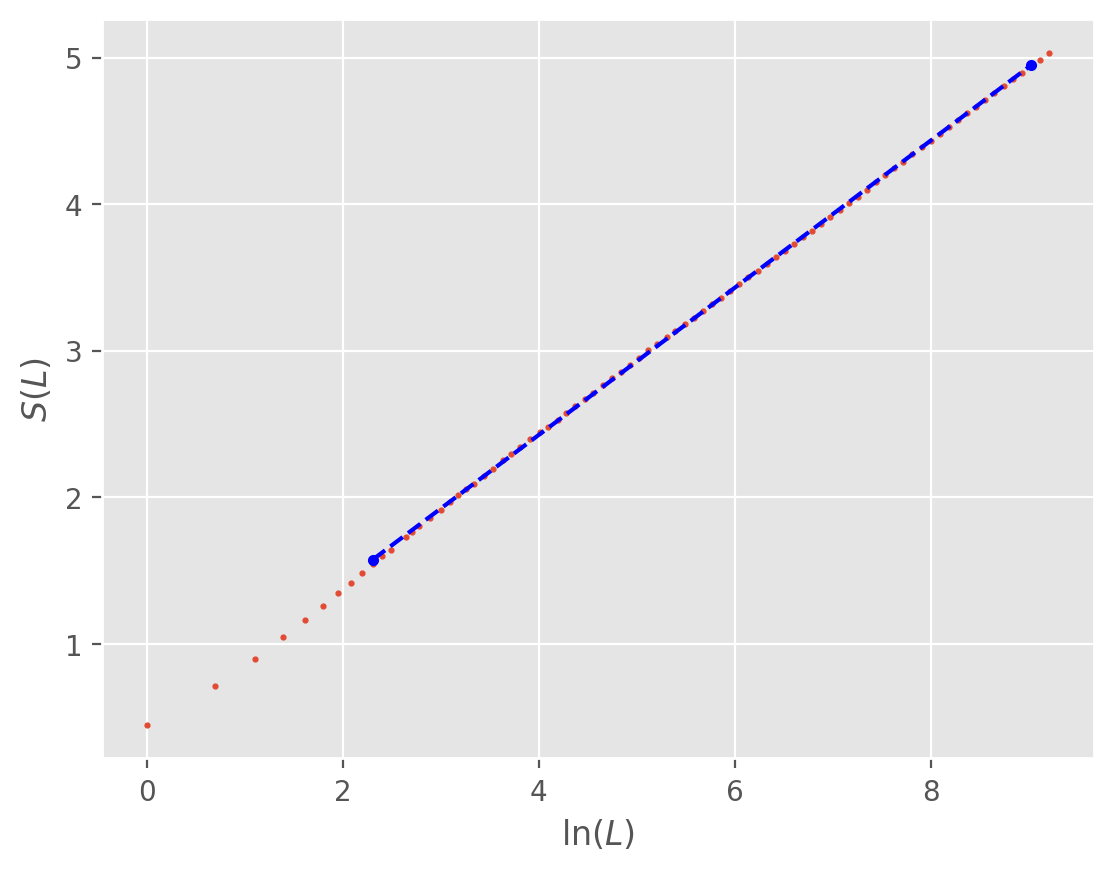}
    \caption{DEA Ito: $\delta = 0.5 \pm 0.0006$}
  \end{subfigure}
  \caption{DEA analysis for both the Stratonovich and Ito versions of the logistic equation.}
  \label{fig:DEA_logistic}
\end{figure}

Our analysis was conducted on ensembles of the logistic equation, and thus, the DEA with stripes utilized a growing window. Both versions of the logistic equation were analyzed. Figure \ref{fig:DEA_logistic} illustrates the results of our analysis near the phase transition $q \approx \alpha$. We believe that our findings shed light on the underlying mechanisms of noise-induced phase transition processes and their impact on extended diffusion processes.
\subsection{Threshold analysis of logistic equation using FPT analysis}
\begin{figure}[ht!]
  \centering
  \begin{subfigure}[b]{0.4\linewidth}
    \includegraphics[width=\linewidth]{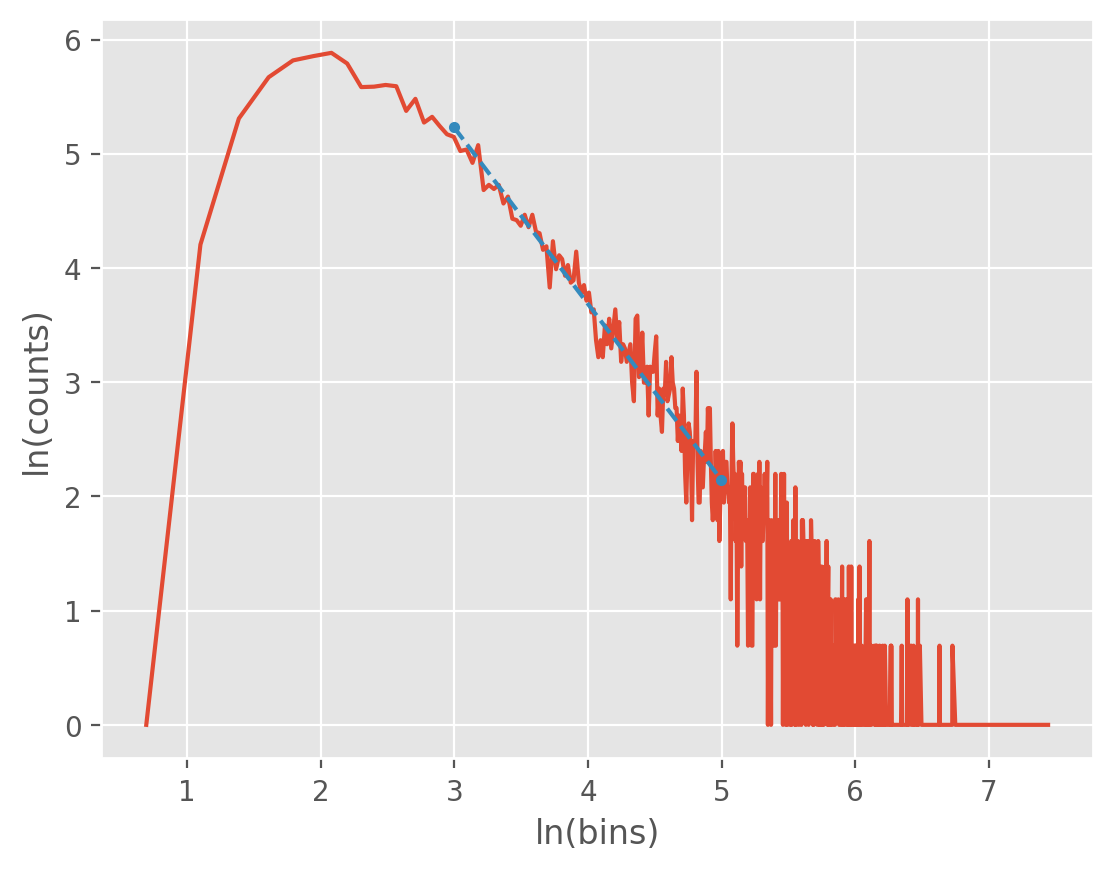}
    \caption{Stratonovich: $ \mu = 1.5 \pm  0.04$}
  \end{subfigure}
  \begin{subfigure}[b]{0.4\linewidth}
    \includegraphics[width=\linewidth]{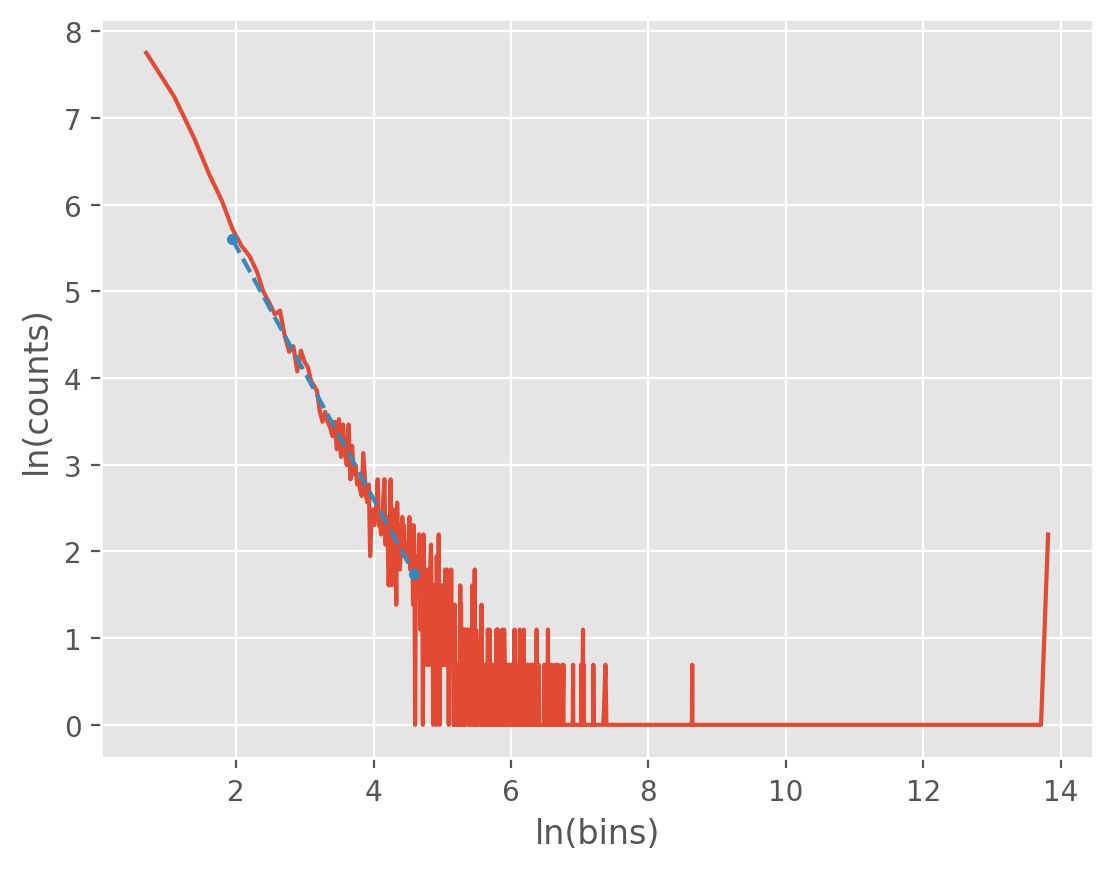}
    \caption{Ito: $\mu = 1.5 \pm  0.004$}
  \end{subfigure}
  \caption{Histograms of first-passage time analysis for both the Stratonovich and Ito versions of the logistic equation. A linear fit was applied to each histogram to calculate $\mu$.}
  \label{fig:FPT}
\end{figure}

In order to further investigate the behavior of the logistic equation in the context of FPT analysis, we employed a threshold method for each version of the equation. Specifically, each series was initialized at $x=0.0001$ and terminated upon crossing the threshold. We then created a log-log plot of the histogram based on the number of steps required for a crossing, as illustrated in Figure \ref{fig:FPT}. To obtain a quantitative measure of the data, we applied a linear fit to determine the slope, denoted by $\mu$. It is worth noting that our numerical results are consistent with the analytical results presented in the previous section.

The FPT analysis approach used here is a powerful tool for characterizing stochastic processes, as it allows us to study the dynamics of systems across time scales. By focusing on the time required for a given system to first reach a specified threshold, we gain insight into the behavior of the system at different levels of complexity. This information can be particularly valuable in understanding the behavior of complex systems that exhibit multiple time scales or other nonlinear dynamics.

Our results demonstrate the utility of the first-passage time analysis approach for studying the logistic equation, and suggest that this approach may be useful for investigating other nonlinear systems as well. Specifically, our findings highlight the potential for this method to uncover underlying trends and dynamics in complex systems, which may not be immediately apparent from other analyses.
\section{Departing from fully random processes in noise-induced phase transitions}
The main focus of our paper is to challenge the traditional form of noise-induced phase transition, as adopted by Schenzle and Brand \cite{schenzlebrand}, by showing that it does not involve the occurrence of CEs. This limitation stems from the assumption that the fluctuation process $\xi$ is fully random, which restricts the model's ability to generate CEs.

To address this limitation, we propose a novel approach based on the quantum-like procedure of Allegrini {et al} \cite{quantum}, which involves generating the fluctuation $\xi$ through a quantum-like operator $\Gamma$ that describes the environment of the variable $x$. Applying the results of \cite{quantum} to the logistic equation of Eq. (\ref{xrepresentation}), we obtain a modified equation that allows for the occurrence of CEs. Specifically, we obtain:

 \begin{eqnarray}\label{crucial}
\frac{\partial }{\partial t}p(x,t)=-\frac{\partial }{\partial x} \left[\alpha x-\beta x^2\right]p(x,t)+q \int_{0}^t dt' <\xi \xi(t')>\frac{\partial }{\partial x}x \frac{\partial }{\partial x}xp(x,t-t'), 
\end{eqnarray}

where $<\xi \xi(t')>$ is a correlation function with the IPL structure necessary to generate CEs. The theoretical prescriptions of \cite{quantum} dictate that the normalized form of this correlation function must be given by 
\begin{equation}
\Phi_{\xi}(t) = \frac{A}{(A^{1/\beta} + t)^{\beta}},
 \end{equation}
 with $0 < \beta < 1$. This function yields CEs with the IPL index $\mu,$ where 
  \begin{equation}
 \mu = 2 + \beta,
 \end{equation}
 and $2 < \mu < 3$, which is in line with the ``intelligence'' of the self-organized processes generating CEs.

We should note that the recent detection of CEs in geophysical fluctuations in California and Hindu Kush \cite{callum} indicates that fault interactions are a process of self-organization generating an IPL index $\mu$, which fits the constraint $2 < \mu <3$. This result supports the 2009 claim \cite{ideal} that the dynamics of the brain, a remarkable example of an ``intelligent" system, host CEs with $\mu$ in the same interval.

In summary, our work provides a novel approach to the logistic equation that allows for the occurrence of CEs, which are crucial for understanding complex self-organized systems. Our approach relies on a quantum-like procedure for generating fluctuations, which has the potential to enhance the model's ability to generate CEs.
\section{Final Remarks}
This study investigated the occurrence of CEs in noise-induced phase transition processes, building on the pioneering work of Schenzle and Brand \cite{schenzlebrand}. We found that the traditional form of these processes does not generate CEs, which aligns with the prevailing belief that a phase transition with $\delta = \frac{1}{2}$ is indicative of a lack of complexity. Instead, we propose that a phase transition with $\delta = \frac{1}{2}$ may be a sign of the existence of pseudo-CEs with an IPL index $\mu = \frac{3}{2}$.

Our findings have important implications for biomedical research, particularly in the fight against cancer growth \cite{medicine}. We suggest that researchers should shift their attention from the traditional approach to noise-induced phase transitions \cite{schenzlebrand} to equations of motion similar to the one proposed in Eq. (\ref{crucial}), which has a time convolution signal indicating the ``intelligence" of the growing system, in this case, cancer. We propose that the key ingredient in preventing the growth of cancer cells is to target the ``errors" that kill the time convolution of the growth process, thereby eliminating the intelligence of the complex system.

Looking ahead, we note that the proposed equation in Eq. (\ref{crucial}) warrants further investigation. In the complexity literature, there is growing interest in explosive phase transition phenomena \cite{important}, particularly in the context of simplicial complexes \cite{boccaletti, ginestra}. The proposal in Eq. (\ref{crucial}) is relevant to the issue raised in \cite{important} on moment closure and other approaches to defining phase transition processes accurately. Additionally, the concept of CEs is related to the phenomenon of Self Organized Temporal Criticality (SOTC) \cite{sotc}, which involves the spontaneous movement of a control parameter to generate criticality. The theory of simplicial complexes aims to evaluate control parameters that generate explosive phase transitions, and our study opens the door to connecting these two seemingly diverging perspectives.

We also note the medical significance of the logistic equation \cite{medicine}. Our proposed modifications to Eq. (\ref{crucial}) establish a connection with the arguments presented in \cite{cousin} on the intelligence of cancer cells that should be targeted for therapeutic purposes. We emphasize that our numerical calculations were performed with an integration time step $\Delta t = 1$, a value much smaller than $1/\alpha = 10$ and $1/\beta = 100$. This choice prevents the emergence of both chaos and the oscillatory dynamics of May, but still allows for the activation of CEs. We propose that the occurrence of neighboring CEs at a time distance $\tau$ close to $1/\alpha$ and $1/\beta$ may establish a connection with the rich complexity originally established by May. This is particularly relevant to the topic of particle swarm optimization, a machine learning approach of increasing interest \cite{pso1,pso2,pso3}.
\section{Declaration of Generative AI and AI-assisted technologies in the writing process}

{\bf Statement:} During the preparation of this work the authors used ChatGPT Mar 14 Version in order to improve the grammar and clarity of the article, and write appropriate highlights and find appropriate keywords. After using this tool/service, the authors reviewed and edited the content as needed and take full responsibility for the content of the publication.
\section*{Declaration of Competing Interests}

The authors declare that they have no conflicts of interest.

\section*{Acknowledgements}
M.B. acknowledges financial support from UTA Mayor project No 8737-22

\bibliographystyle{elsarticle-num}

\end{document}